# Simulation of a particle domain in a continuum/fluctuating hydrodynamics reservoir

Abbas Gholami,[1, *] Rupert Klein,[1, †] and Luigi Delle Site[1, ‡]

[1]*Freie Universität Berlin, Institute of Mathematics, Arnimallee 6, 14195 Berlin, Germany*

In molecular simulation and fluid mechanics, the coupling of a particle domain with a continuum representation of its embedding environment is an ongoing challenge. In this work, we show a novel approach where the latest version of the adaptive resolution scheme (AdResS), with non-interacting tracers as particles' reservoir, is combined with a fluctuating hydrodynamics (FHD) solver. The resulting algorithm, supported by a solid mathematical model, allows for a physically consistent exchange of matter and energy between the particle domain and its fluctuating continuum reservoir. Numerical tests are performed to show the validity of the algorithm. Differently from previous algorithms of the same kind, the current approach allows for simulations where, in addition to density fluctuations, also thermal fluctuations can be accounted for, thus large complex molecular systems, as for example hydrated biological membranes in a thermal field, can now be efficiently treated.

## INTRODUCTION

Molecular Dynamics (MD) [1] is considered a valuable tool for studying molecular systems at the microscopic level. Its contribution in condensed matter and molecular science covers many subjects of high interest in current research, such as molecular docking [2] or the design of materials for advanced technologies [3].

An MD model simulates the motion of molecules according to Newton's equations of motion. Molecules, in the most popular approach, have atomistic resolution according to their chemical structure. That is, they are represented by spherical atoms kept together by rigid bond constraints or intramolecular potentials, [4]. The interaction between molecules is governed by potentials acting between their respective atoms, also called force-fields. The latter are derived from experimental data or from elaborate electronic structure calculations (see, e.g., [5] for the important case of liquid water).

The core of the technique is thus the time integration of a highly complex Hamiltonian system with a state space given by the collection of the positions and momenta, $(\mathbf{r}, \mathbf{p})$, of all the atoms considered. Assuming ergodicity of this system, the effective value of the resulting simulations lies in their capability of potentially sampling, through long enough trajectories, classical statistical ensembles and thus allowing the user to study the system's macroscopic thermodynamic properties. More precisely, one exploits the insight that under the ergodic hypothesis and considering large time windows, the time of residence of the system in regions of the state space with given total energies of their microstates is proportional to the volume of these regions [1].

The consequence is that through the instantaneous value of a physical observable, $w(\mathbf{r}, \mathbf{p})$, its average, $W$, can be calculated over a sufficiently large time window along the trajectory of the MD simulation, that is: $W = \langle w(\mathbf{r}, \mathbf{p}) \rangle_{\text{time}} = \lim_{t \to \infty} \int_0^t \frac{1}{t'} w(\mathbf{r}(t'), \mathbf{p}(t'), t') \, dt'$. Specific thermodynamics and statistical mechanics ensembles are realized in this context by isolating the system entirely from its surroundings (microcanonical), bringing it in contact with the heat source/sink of an external thermostat (canonical), or allowing for the exchange of heat and particles with a reservoir (grand canonical), [1, 6]. In this fashion, the atomistic simulations allow for the microscopic/particle-based analysis of the statistical collective behaviour of large molecular systems. A remarkable example is liquid water for which an atomistically resolved simulation enables detailed analyses of the statistical properties of, e.g., the hydrogen bonding network and its crucial role in determining the multifaceted behaviour of water as a solvent in both nature and technological systems.

With the increase of the system size computational cost rise steeply, however, so that macroscopic simulations at atomistic resolution become prohibitively expensive. Therefore, the simulation of complex molecular systems requires efficient computational strategies that capture the essence of a physical/chemical process at reasonable computational costs. In particular, for the large class of problems with the event of interest occurring merely in a limited region of space, it is convenient to reduce the simulation to a relatively small high-resolution region represented with atomistic detail, coupled to an effective simplified environment that plays the role of a macroscopic thermodynamic bath. The challenging aspect of such a simplification is the definition of boundary conditions between the particle domain and the environment that assures the exchange of energy and matter consistently with the fully microscopically resolved system of reference.

Several approaches have been proposed during the last decade under the umbrella of the adaptive resolution technique or similar models [7–11]. In particular, the last AdResS version, developed by the authors [12–15] has been framed into a more general mathematical model of open systems which assures, in a systematic fashion, the statistical mechanics' consistency of the high-resolution region with respect to an open system embedded in a fully microscopically resolved environment [16, 17]. The supporting mathematical model was also instrumental to the definition of boundary conditions for situations beyond equilibrium and was successfully tested to study open molecular systems in a stationary thermal gradient [18, 19].

Current molecular science moreover demands to go even beyond constant thermal or density fields and treats molecular systems embedded in arbitrarily fluctuating fields. For instance, cell membranes in a realistic environment are subject to a fluctuating thermal field that has a major impact on their hydration properties and morphological structure [20, 21]. In technology, the possibility of externally changing/modulating in time a thermal or density field could be used to build efficient devices, e.g., for phase separation of liquids in the context of water purification [22, 23].

A large-scale fluctuating environment can be described efficiently by continuum mechanics without the need for atomistic resolution. Macroscopic physical quantities in the form of space and time-dependent fields then describe the collective behaviour of the particle system. The Navier-Stokes equations constitute the model of choice for liquids in this context. They describe the dynamics of a viscous fluid based on the conservation laws for mass, momentum, and energy of a Newtonian fluid [24]. Thus, the fluid mass within an arbitrary control volume can only change by exchange of mass with its environment but not by the action of sources or sinks. The same holds for the conservation of momentum with the total momentum exchange being due to the advective exchange of momentum-carrying mass elements and due to the forces acting on the surface of the volume by the pressure and viscous stress fields. Finally, energy is exchanged again by the advection of energy-carrying mass elements, by the work of the pressure and viscous forces on the element, and by thermal energy transport through heat conduction. Neglecting gravity, the Navier-Stokes equations, written in conservation form, read as

$$\frac{\partial \varrho}{\partial t} + \nabla \cdot (\varrho \mathbf{u}) = 0 \qquad (1a)$$

$$\frac{\partial \varrho \mathbf{u}}{\partial t} + \nabla \cdot (\varrho \mathbf{u} \circ \mathbf{u} + p \operatorname{Id} + \boldsymbol{\tau}) = 0 \qquad (1b)$$

$$\frac{\partial \varrho e}{\partial t} + \nabla \cdot ([(\varrho e + p)\operatorname{Id} + \boldsymbol{\tau}]\mathbf{u} + \mathbf{j}) = 0, \qquad (1c)$$

where $(\varrho, \mathbf{u}, e)$ are the density, flow velocity, and specific internal energy, respectively, and Id is the unit tensor. The thermodynamic pressure $p$, the viscous stress tensor $\boldsymbol{\tau}$, and the heat flux density $\mathbf{j}$ are given by the fluid's constitutive laws

$$p = \widetilde{p}(\varrho, \mathbf{u}, e) \qquad (2a)$$

$$\boldsymbol{\tau} = -\nabla[\mu(\nabla \mathbf{u} + (\nabla \mathbf{u})^\top)] + \nabla[\lambda(\nabla . \mathbf{u})\mathbf{I}], \qquad (2b)$$

$$\mathbf{j} = -k\nabla T, \qquad (2c)$$

where $T = T(\varrho, p)$ is the temperature, and where we have explicitly adopted Newtonian friction in the equation of state for the stress tensor, with $\mu$ and $\lambda$ being the shear and volume viscosities, respectively, and $k$ the heat conductivity. Solving Navier-Stokes equations (Eqs.1) together with the equations of state (Eqs.2) will determine the flow behaviour in the system for density, velocity, pressure, and temperature fields.

As they are formulated in Eqs.1-2, the continuum equations hold in the limit of asymptotic scale separation between the atomistic and the systems scales. Of interest here, however, are large-scale flows on mesoscale for which the continuum dynamics still involves sizeable thermal fluctuations. To describe this situation, Landau and Lifshitz pioneered the formulation of a (linear) "fluctuating hydrodynamics" model [25], which essentially extends the deterministic conservation laws from Eqs.1-2 by stochastically fluctuating flux densities for mass, momentum, and energy. See reports on further extensions of this theory and on computational implementations, e.g., in [26, 27]. In this paper, we describe a hybrid computational methodology that allows us to pursue molecular simulations at atomistic detail in limited regions that are embedded in mesoscale environments with general space- and time-dependent statistical evolution. That is, we describe a new variant of the AdResS technology that enables the systematic coupling of an open molecular system to a mesoscale environment governed by fluctuating hydrodynamics.

The examples discussed above lead to the simulation/modelling question of how to define boundary conditions that couple the particle domain to a fluctuating environment described by a much less detailed mathematical model. Such boundary conditions need to be designed so that the statistical and thermodynamic conditions of the particle domain adapt to the instantaneous information coming from the fluctuating environment and, vice versa, the fluctuating environment adapts to the instantaneous response of the particle domain.

In this work, we realize such an algorithm by coupling the AdResS scheme (for the open molecular system) to a continuum fluctuating hydrodynamics scheme based on the Navier-Stokes model (for a fluctuating thermodynamic environment) via boundary conditions inspired by the mathematical model of the open system. The method is tested for fluid Argon under different initial conditions, following its relaxation to equilibrium. The relaxation is specifically analyzed in the particle domain of AdResS and shows that it leads to the expected behaviour from a continuum as well as from the corresponding fully resolved particle point of view. It must be reported that previous attempts of coupling particle domains to the continuum and in particular with other versions of AdResS have been technically satisfactory for the description of density fluctuations [27–32]. However, the current work with a systematic one-to-one correspondence between the technical implementation and the mathematical and physical formulation of boundary con-ditions at the microscopic level produced an algorithm with a major improvement, i.e., it allows, in addition to density fluctuations, for thermal fluctuations whose description is not possible in previous algorithms.

## ADRESS VS. THE MATHEMATICAL OPEN SYSTEM MODEL

The latest version of the AdResS method directly couples a high resolution (atomistic) region of physical interest, AT, to a region of non-interacting point particles (tracers, TR) through an interface region $\Delta$ at atomistic resolution. In the AT and $\Delta$ regions particles undergo standard atomistic interactions as in a full atomistic simulation. In the $\Delta$ region they are, in addition, subject to a thermostat and to an external and purely space-dependent one-particle force called the "thermodynamic force", $F_{th}(x)$. Such force assures, together with the action of the thermostat, the thermodynamic consistency of the $\Delta$ and AT regions. Specifically, the thermodynamic force is calculated self-consistently during an equilibration run of AdResS: Starting from a first guess $F_{th}^{(0)}(x) = 0$, the update at step $k$ is $F_{th}^{(k+1)}(x) = F_{th}^{(k)}(x) - c\nabla \varrho_k(x)$, with the density profile $\varrho_k(x)$ calculated from an AdResS simulation using $F_{th}^{(k)}(x)$; $c > 0$ is a coefficient that controls the speed of convergence. The iteration stops when the deviation of $\varrho_k(x)$ from a reference density profile is within a prescribed tolerance. Once $F_{th}(x)$ has been determined, it remains unchanged without any need for recalibration in subsequent applications [33]. A statistical mechanics analysis of this set-up allows one to identify $F_{th}(x)$ with the correction needed to balance the chemical potential of the TR+$\Delta$ region to the chemical potential of reference the AT region [14, 34, 35]. In addition, the thermostat assures the thermal consistency of the $\Delta$ and AT regions at the target temperature. The tracer region, TR, acts as an artificial reservoir of particles; the entrance/exit of particles into/from the $\Delta$ region is regulated by the thermodynamic force and assures the proper balance (see Fig.1).

This scheme has been proved to reproduce the features of a Grand Canonical ensemble [15, 35]. The numerical setup, in turn, inspired the development of a physico-mathematical model of the open system that provided further conceptual support to the definition of AdResS as a numerical method to simulate open systems [16, 17]. Specifically, the mathematical model concludes that to properly simulate an open system, it is sufficient to define the boundary conditions in $\Delta$ without the need of requiring a physically meaningful behaviour of the particles in the TR region. The analytic details of the physico-mathematical model can be found in Refs.[16, 17], here we provide just a short, albeit essential, summary:

We consider a large, closed, system with $N$ interacting particles named "Universe" and the corresponding Liouville equation for its phase-space probability distribution $F_N(\mathbf{X}, t)$; $\mathbf{X} \equiv \{\mathbf{X}_1, ....\mathbf{X}_N\}$; $\mathbf{X}_i \equiv (\mathbf{q}_i, \mathbf{p}_i)$; $(i = 1, ..., N)$ where $\mathbf{q}_i$ and $\mathbf{p}_i$ are the position and momentum of the $i$-th particle and $t$ is the time. Let us now assume our main interest is in some open subsystem, $\Omega$, of the Universe that contains a time-dependent instantaneous number of particles $n$. The equivalent of the Liouville equation for $\Omega$ results in a hierarchy of equations for the probability distribution functions in the phase-spaces of $0 \leqslant n \leqslant N$ particles in $\Omega$, labelled $f_n(\mathbf{X}_\Omega, t)$. The equation for the evolution of $f_n(\mathbf{X}_\Omega, t)$ is derived by integrating all the particles' degrees of freedom outside $\Omega$ in the Liouville equation of the Universe. Differently from the Liouville equation of the Universe, the equation for $f_n(\mathbf{X}_\Omega, t)$ is characterized by a term describing the coupling of $\Omega$ to the external reservoir. Such a term implies that to have a physically consistent exchange of energy and particles between $\Omega$ and the reservoir, particles at the boundary, exiting from or entering in $\Omega$, should be distributed according to the one- and two-particle distribution function of reference (i.e., as in the fully resolved Universe) at the given temperature. Under conditions of equilibrium, such a set-up is shown to be consistent with a Grand Canonical representation of $\Omega$.

The close similarity of this model to the computational set-up of AdResS lies in the fact that the abovementioned conditions regarding the information exchange between the open system and the surrounding reservoir(s) are effectively implemented in AdResS through the action of the thermodynamic force and the thermostat in $\Delta$. In fact, at the boundary of the AT region the targeted tem-perature and density/one-particle distribution function are imposed while the radial (two-particle) distribution function is automatically recovered and used as a criterion of validation of AdResS (see also Refs. [13, 17]). Furthermore, when an open system is in contact with two distinct reservoirs, the mathematical model suggests that the coupling at each system-reservoir boundary should be done as if the system was in equilibrium with the single reservoir, independently from the other. In AdResS, the corresponding effective numerical condition consists in applying at each boundary of AT the thermodynamic force and the thermostat corresponding to the thermodynamic state of each reservoir [17, 19].

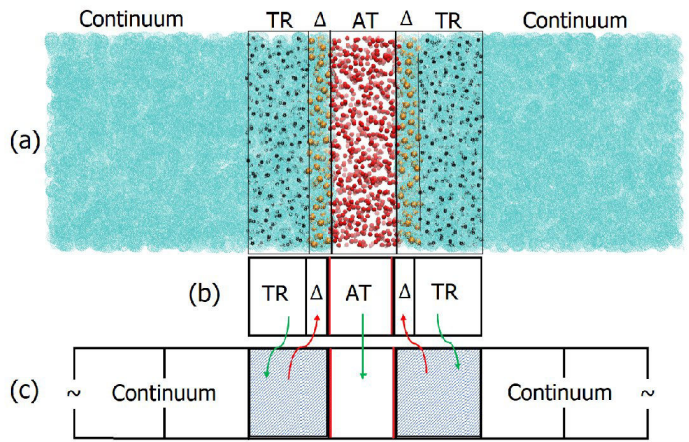

Figure 1. Illustration of the AdResS/open system model and its coupling to the continuum (a), the technical set-up of AdResS only (b), and its inclusion into a continuum solver as particle domain (c). The particles of the $\Delta$ region and the tracers of the $TR$ region of AdResS overlap with the continuum description and exchange information (green and red arrows) in a way that boundary conditions from the continuum to the AT region and vice versa are defined.

This framework, considering explicitly the time dependence of the whole system, allows for a further relevant step forward, that is to account for reservoirs with fluctuating temperature and density. According to our model, the corresponding boundary condition must allow for the instantaneous exchange of particles and energy according to the time-dependent thermodynamic conditions at the system-reservoir interface. Accordingly, in AdResS, the effective realization of the idea above consists of applying at time $t$ a thermodynamic force in $\Delta$ that corresponds to the instantaneous thermodynamic condition of the reservoir. The technical essence of the present work is the numerical implementation of this concept, with the thermodynamically fluctuating reservoir implemented via fluctuating hydrodynamics (FHD) and synchronized with the particle system through a varying thermodynamic force tabulated in a predefined dictionary of thermodynamic forces. Below the scheme is explained in detail; instead, the explicit connection to the mathematical model, is reviewed in the supporting material which is complemented by Refs.[14, 16–18, 34].

## COUPLING ADRESS AND FHD

The description of the macroscopic reservoir, as anticipated above, is achieved through the FHD model, that is in essence Navier-Stokes (NavSt) equations extended by the addition of a stochastic flux term. In fact, in statistical mechanics, fluctuations are random deviations of a system from its average state as the system does not stay at the microscopic state of equilibrium but randomly samples all possible states with a Boltzmann distribution probability [36, 37].

To incorporate fluctuations into macroscopic deterministic hydrodynamics, Landau and Lifshitz introduced an extended form of the NavSt equations by incorporating a stochastic flux divergence term, $\mathbf{S} = (0, \mathcal{S}, \mathcal{Q} + \mathbf{u} \cdot \mathcal{S})^\intercal$ [25]. The Landau-Lifshitz Navier-Stokes equations are written as: $\mathbf{U}_t = -\nabla \cdot (\mathbf{F} + \mathbf{D} + \mathbf{S})$, where $\mathbf{U} = (\varrho, \mathbf{J}, E)^\intercal$ is a vector of conserved quantities with $\varrho$, $\mathbf{J} = \varrho\mathbf{u}$, and $E = \varrho e$ being the mass, momentum, and energy densities, respectively. The advective ($\mathbf{F}$) and diffusive ($\mathbf{D}$) fluxes are given by: $\mathbf{F} = (\varrho\mathbf{u}, \varrho\mathbf{u}\cdot\mathbf{u}, (E+P)\mathbf{u})^\intercal$ and $\mathbf{D} = (0, \tau, \tau\cdot\mathbf{u} - k\nabla T)^\intercal$; where $\mathbf{u}$ is the velocity, $P$ is the pressure, $T$ is the temperature, and $\tau = -\eta(\nabla\mathbf{u} + (\nabla\mathbf{u})^\intercal - \frac{2}{3}I\nabla\cdot\mathbf{u})$ is the stress tensor in which $\eta$ and $k$ are the coefficients of viscosity and thermal conductivity, respectively. The stochastic stress tensor ($\mathcal{S}$) and heat flux ($\mathcal{Q}$) are white in space and time and are formulated using fluctuation-dissipation relations to yield the equilibrium covariances of the fluctuations with the mean value of zero and well-specified covariances [26]. Different discretization techniques are available for solving such equations as reported in the supplementary material which is complemented by Refs.[26, 38–42].

As a technical reference for the coupling of a particle system to FHD, we utilized the state-flux hybrid scheme of Donev and coworkers [39]. In their technique, the output of the continuum solver at the neighbouring cells of the particle subdomain acts as a boundary condition for the state of the particle subdomain. Particles with specific positions and velocities are inserted into the reservoir to reproduce the mass, momentum, and energy densities of the corresponding continuum cells. Conversely, the total mass, momentum, and energy of particles crossing the boundary and passing from the particle subdomain to the continuum reservoir are calculated during the particle simulation and implemented as a flux boundary condition for the continuum solver at the interface.

The fluctuating thermodynamic conditions of the FHD-simulated reservoir are accounted for in the AdResS-system by utilizing a dictionary of thermodynamic forces in the $\Delta$ region that has been pre-calculated as a function of density and temperature. The states obtained by the continuum solver in the grid cells immediately adjacent to the particle subdomain define the respective thermodynamic forces to be used in the next time step for the particle-based solver. In the process, parameters of the thermodynamic force functions are interpolated between dictionary entries where needed. In the other coupling direction, the calculation of the fluxes in the particle domain to be imposed to the continuum solver as a boundary condition is done as in the reference technique Ref. [39]. That is, the total mass, momentum, and energy of particles crossing the particle-continuum interface is imposed as a boundary condition to the continuum solver at the interface. The calculated flux values are consistent with the conservation laws of mass, momentum, and energy (see also the supporting material which includes Refs.[43–45]). The resulting scheme is summarized in Fig.2 and its numerical validation is discussed in the Appendix. Another non-trivial further technical advancement compared to previous schemes [29, 46] is that the current algorithm does not require an additional optimization step upon the insertion of particles from the continuum region. In fact, the tracer particles are non-interacting objects and their entrance in the $\Delta$ region, as well as their subsequent equilibration with the local environment, are automatically regulated by the thermodynamic force within the $\Delta$ region.

## CONCLUSIONS

A novel approach for simulating an open system at particle resolution embedded in a reservoir of energy and particles modelled by fluctuating hydrodynamics (FHD) has been presented. The mathematical model at the core of the AdResS algorithm allows prescribing coupling conditions that smoothly allow for an automatic dynamical exchange of particles and energy. The AdResS approach has already been demonstrated to be capable of representing complex molecular systems in and out of equilibrium by coupling it to several reservoirs at different thermodynamic states along the surface of the AT region. Thus, the new description of adjacent reservoirs through FHD allows for the simulation of similar systems but this time in the presence of fluctuating thermal and/or density fields. The option of coupling an AT region to several adjacent reservoirs in combination with our non-stationary FHD extension also constitutes a promising basis for multi-dimensional variants of AdResS-FHD systems, including situations with spatially inhomogeneous state distributions along a (flat) AdResS surface. Such situations would be covered by treating the faces of all FHD cells that bound on the AT interface in question as separate reservoirs. The linearity of the coupling to several reservoirs, demonstrated in previous work [17], will make this straightforward to implement.

## ACKNOWLEDGMENTS

This research has been funded by Deutsche Forschungsgemeinschaft (DFG) through grant CRC 1114 "Scaling Cascade in Complex Systems", Project Number 235221301, Project C01 "Adaptive coupling of scales in molecular dynamics and beyond to fluid dynamics". The authors thank Felix Höfling and Roya Ebrahimi Viand for the helpful discussion and support on using the HALMD package.

## APPENDIX ON THE NUMERICAL SCHEME AND ITS VALIDATION

Figure 2 reports the schematic structure of the numerical scheme; here, we discuss the results of the numerical simulation that validates the method. Several numerical tests for a one-dimensional coupling set-up are reported upon here. Thus, the FHD-to-AdResS change of resolution occurs only along one dimension. The coupling between the computational system components follows the principles outlined above. Additional technical details can be found in the supporting material which includes Refs.[47–52]. A one-dimensional coupling set-up is simple enough to allow for simulations that clearly assess the general validity of the basic principles on which the technique is based and at the same time it is already sufficient for applications to complex molecular systems. One concrete example, previously mentioned, concerns how a temperature field (gradient) affects the geometry of biological membranes. The hypothesis is that even small temperature variations across the membrane could generate unexpected shape responses leading to the conclusion that the shape response of a membrane can be tuned by externally controlling a temperature gradient in its immediate vicinity at the two different sides of the membrane [53]. A prototype of the hydrated membrane, in absence of thermal gradient, has already been successfully treated with the AdResS technique [54], thus the current approach can now make the crucial further step forward by adding the thermal fluctuations along the direction that crosses the membrane (that is in a one-dimensional coupling set-up). Another relevant example, also mentioned before, concerns water-ionic liquid mixtures. In such systems, a tunable one-dimensional thermal field can drive a phase separation in water-rich and water-poor domains. Also, in this case, the AdResS simulation has been successfully applied in absence of external gradients [23, 55] and is now ready for the next step offered by the current algorithm.

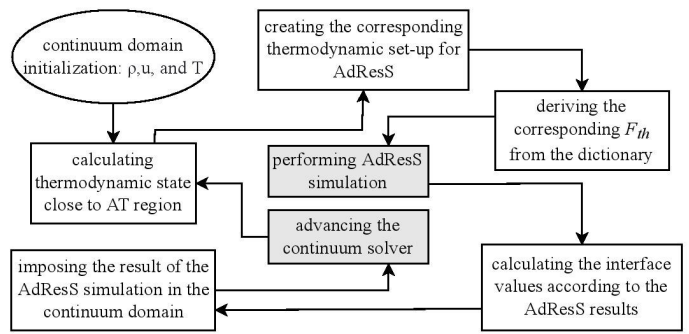

Figure 2. The definitions of boundaries of Fig. 1 are considered. The simulation starts with an initial state of a given density, velocity, and temperature for the whole domain as if it was a pure continuum domain. Next, according to the density and temperature of the neighbouring macrocells of the continuum, the thermodynamic forces for the left and right sides of the AdResS domain are calculated by interpolation from the dictionary and the particle simulation is executed. The temperature on the left and right sides of the particle domain is set according to the temperature of the neighbouring macrocells. Average values of physical quantities of interest and the related interface fluxes are calculated from the particle simulation in the AT region and imposed in the continuum macrocells corresponding to the AT region and in the neighbouring/interface cells to ensure that such imposing preserve the conservation laws. Finally, the fluctuating hydrodynamic solver will advance for a certain number of continuum time steps, considered as a single coupling time step, and a new state with updated density, velocities, and temperature is defined and the procedure repeated.

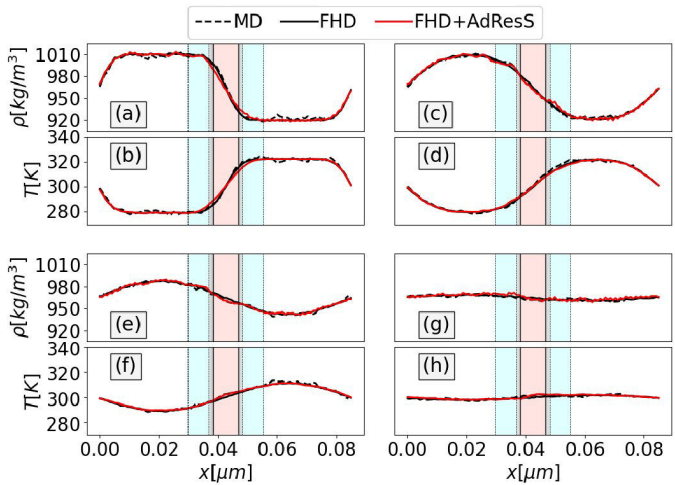

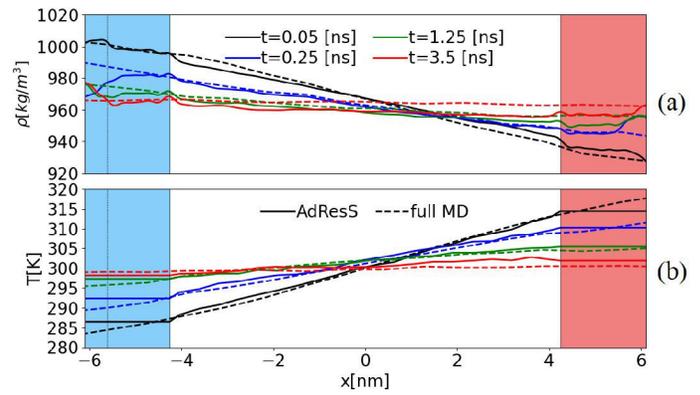

Figure 3. Profiles of density (a, c, e, g) and temperature (b, d, f, h) evolved from initial step functions ($\varrho_{left} = 1010[kg/m^3]$, $T_{left} = 279.3[K]$, $\varrho_{right} = 919.6[kg/m^3]$, and $T_{right} = 322.3[K]$) with constant pressure ($P = 100[MPa]$) over time at $t = 0.05[ns]$ (a and b), $t = 0.25[ns]$ (c and d), $t = 1.2[ns]$ (e and f), and $t = 3.5[ns]$ (g and h). The colored regions in the middle of the box show the AdResS domain with the atomistic region (red) in contact with the TR region (blue) through a narrow transition region (gray). Results from FHD, MD and coupled AdResS-FHD simulations are shown as black solid, dashed and red lines, respectively.

Figure 4. The profile of density (a) and temperature (b) in the AdResS domain and in the full atomistic simulation over time, for the initial step function for the density and temperature. The red and blue regions represent the hot and cold reservoirs, respectively. The time slices are the same as for Fig.3.

As numerical tests, we run simulations in each of which a specific initial condition for density, velocity, temperature, and pressure is given. We then follow the evolution of the system and compare the result of the AdResS-FHD algorithm with the results of the full continuum and the full atomistic simulations. As a representative of all the tests, Fig. 3 reports the results obtained for a Riemann initial value problem given by zero velocity, constant pressure, and piecewise constant initial data for the density (and temperature) with the discontinuity located in the center of the domain. We follow the relaxation in time to equilibrium and show that indeed the AdResS-FHD simulation behaves as expected from the full FHD and full MD simulations. In particular, in the AT region of AdResS, we find that the evolution towards equilibrium for the aforementioned test scenario occurs as in the equivalent subsystem of a full atomistic simulation of the entire domain (see Fig. 4). The results are of similar quality when instead of said initial contact discontinuity an acoustic wave with periodic initial conditions for density and temperature is imposed, but this time with a non-flat initial pressure profile consistent according to the equation of state of the system. Such results are reported in the supporting material.

In the supporting material (including Refs.[56–58]), we also report the case where the system is treated in a quasi one-dimensional set-up with varying cross-section geometry. Results show that indeed the AdResS-FHD algorithm satisfactorily reproduces the behaviour of reference.

---

* agholami@zedat.fu-berlin.de
† rupert.klein@fu-berlin.de
‡ luigi.dellesite@fu-berlin.de

# Supplementary Material on "Simulation of a particle domain in a continuum/fluctuating hydrodynamics reservoir"

Abbas Gholami,[1] Rupert Klein,[1] and Luigi Delle Site[1]

[1]*Freie Universität Berlin, Institute of Mathematics, Arnimallee 6, 14195 Berlin, Germany*

## ADAPTIVE RESOLUTION SIMULATION (ADRESS)

### Coupling term of the mathematical model and the AdResS algorithm

In this section, we review the connection between the system-reservoir coupling term of the mathematical model and the corresponding computational conditions. In Ref.[1], the mathematical conditions of this connection are derived in detail. Here, instead we highlight the relevance of the time-dependent framework for simulating thermodynamic fluctuations of the reservoir(s), which is presented, in its essence, in the main paper. The system-reservoir coupling term obtained by integrating out the degrees of freedom of the particles outside the open system in the Liouville equation of the entire system consists of two components:

$$\Psi_n = -\sum_{i=1}^{n} \nabla_{\mathbf{p}_i} \cdot \left( \mathbf{F}_{\mathrm{av}}(\mathbf{q}_i) f_n(t, \mathbf{X}^{i-1}, X_i, \mathbf{X}_i^{n-i}) \right) \quad (1)$$

where $\mathbf{F}_{\mathrm{av}}(\mathbf{q}_i) = -\int_{S^c} \nabla_{\mathbf{q}_i} V(\mathbf{q}_i - \mathbf{q}_j) f_2^\circ(X_j|X_i) dX_j$, is the mean-field force exerted by the outer particles (with phase-space domain $S^c$) onto the $i$th inner particle, with $f_2^\circ(t, X_{\mathrm{out}}|X_{\mathrm{in}})$ being the conditional distribution for joint appearances of an outer particle given the state of an inner one, this quantity is unknown (since the degrees of freedom of the reservoir have been integrated out) and needs to be assumed/modeled. The second term is:

$$\Phi_n^{n+1} = (n+1) \int_{\partial\Omega} \int_{(\mathbf{p}_i \cdot \mathbf{n})>0} (\mathbf{p}_i \cdot \mathbf{n}) \left( f_{n+1}(t, \mathbf{X}^n, (\mathbf{q}_i, \mathbf{p}_i)) - f_n(t, \mathbf{X}^n) f_1^\circ(t, \mathbf{q}_i, -\mathbf{p}_i) \right) d^3 p_i d\sigma_i \quad (2)$$

where $f_{n+1}(t, \mathbf{X}^n, (\mathbf{q}_i, \mathbf{p}_i))$ is the probability distribution of the open system when particles are $n+1$ while $f_1^\circ(t, \mathbf{q}_i, -\mathbf{p}_i)$ is the one-particle distribution of the reservoir at the interface boundary $\partial\Omega$. As for $f_2^\circ$, also $f_1^\circ$ is unknown and needs to be assumed/modeled. Thus, if one wishes to have an open system with physical consistency, the mandatory condition is to model $f_1^\circ$ and $f_2^\circ$ in a realistic manner. The most realistic model of such quantities is the corresponding quantities calculated in a fully resolved system of which the open system is a subsystem.

In terms of the coupling algorithm, the two conditions above mean that at the interface between the AT region and the $\Delta$ region one needs to make the assumption for $f_1^\circ(t, \mathbf{q}_i, -\mathbf{p}_i)$ and $f_2^\circ(t, X_{\mathrm{out}}|X_{\mathrm{in}})$ which are the unknown quantities of the problem. In fact, all the other quantities involved in Eqs.1 and 2 are explicitly or implicitly provided by the molecules of the AT region which are explicitly simulated. To this aim, in AdResS, the $\Delta$ region has been defined as an interface region where the thermodynamic force imposes a one-particle density equivalent to the one-particle density that the full system would have if it was entirely resolved with the atomistic resolution, while the thermostat fixes the temperature at the target value (thus implicitly targeting the distribution of momenta). In such a manner, $f_1^\circ$ in AdResS is fixed as it was in a full resolved system. Regarding $f_2^\circ$, in principle one can also devise a procedure of calculation as for the thermodynamic force currently used; however, it has been found numerically that the imposition of $f_1^\circ$ leads to a two-body distribution function in AT and $\Delta$ which closely reproduce the two-body distribution function of a fully resolved system, thus the imposition of $f_1^\circ$ is sufficient to numerically model both unknown quantities in a satisfactory manner [2]. For stationary/equilibrium situations, the same conditions can be derived from a Grand Canonical-like analysis of the open system [3, 4]; however, such an approach is no more justified in the case of fluctuating thermodynamics conditions. Here lies the novelty of the mathematical model of Ref.[5] that we have discussed in this paper because the formalism of the Liouville equation considers generic time-dependent particle distribution functions and thus allows for the inclusion of fluctuating thermodynamic conditions in the system. The extension of the algorithm done in this paper goes beyond the use of the Grand Canonical-like model as a theoretical reference and extends the boundary conditions to time-dependent boundary conditions through the dictionary of thermodynamic forces which respond to the thermodynamic fluctuations coming from the reservoir.

### Technical details

We have considered Argon as a single-atomic fluid with the relative simplicity of the spherical two-body pair potential and compliance with the basic theory. In all MD simulations, the simulation box is filled with Lennard-Jones (LJ) particles with Argon's length and energy pa-rameters of $\sigma = 0.34[nm]$ and $\varepsilon = 120k_\mathrm{B}[J]$ where $k_\mathrm{B}$ is the Boltzmann constant. In all cases, AdResS simulation box is partitioned as: $AT = 25\sigma$, $\Delta = 4\sigma$, and $TR = 21\sigma$. Particles in $\Delta$ and AT regions interact with LJ potential reads as $U_{LJ} = 4\varepsilon[(r/\sigma)^{-12} - (r/\sigma)^{-6}]$ with $r$ being the inter-particle distance. The inter-particle potential is set to zero at distances larger than the cut-off radius of $r_c = 2.5\sigma$.

In AdResS with the abrupt change of resolutions, particles may experience unphysical large forces while entering the $\Delta$ region from the TR region as the inter-particle interaction is on and off in $\Delta$ and TR regions, respectively. In such scenarios, we cut those single particles' forces to the average inter-particle forces to prevent the failure of simulations. However, it has been shown that such correction will not affect the dynamics of the region of interest (AT)[3]. The threshold for capping forces at the interface of $\Delta$ and TR regions is set to $500\varepsilon/\sigma$ for each Cartesian component of the force.

We have considered a supercritical state far from the critical point and coexistence region at the reduced temperature and density of $T^* := k_\mathrm{B}T/\varepsilon = 2.5$ and $\varrho^* := \varrho\sigma^3 = 0.57$ corresponding to a reduced pressure of $P^* := P\sigma^3/\varepsilon = 2.37$. These reduced values correspond to the following SI units for Argon: $\varrho = 964.82[kg/m^3]$, $T = 300[K]$, and $P = 100[MPa]$. This density means having 9647 Argon particles in the AdResS simulation box of size $75\sigma \times 15\sigma \times 15\sigma$. The temperature and density of the AdResS simulation while coupling to FHD depend on the thermodynamic state of the simulation at each time step. However, for the equilibrium case, the thermodynamic state point is considered the abovementioned supercritical state. For the calculation of the thermodynamic forces, the criterion for stopping the iterative procedure is set to $max|\varrho(x) - \varrho_t|/\varrho_t < 1.5\%$ where $\varrho_t$ is the target flat density and the maximum is taken across the whole AdResS simulation box. Usually, 10-20 iterative steps are required to find the proper thermodynamic force. For calculating the density profile as a function of length by Fourier transformation, the box is divided into 750 slices along the axis of change of resolution.

All MD simulations were carried out with the HAL's MD package[6, 7] and H5MD format of input and output files [8] with a Verlet integrator with the time step of $0.002\tau$ in which $\tau$ is the reduced unit of time $\tau := \sqrt{m\sigma^2/\varepsilon}$ with $m$ being the mass of Argon atoms ($6.6335209 \times 10^{-26}[kg]$); thus, the MD time step in SI units is $4.3[fs]$. We have used the Andersen thermostat[9] with the update rate of $\nu = 15/\tau$ which means particles' velocity is re-sampled every 33 integration steps from the Maxwell-Boltzmann distribution. The relaxation time at each step for finding the thermodynamic force is $5000\tau$ while the initial 30% of simulation time is discarded and considered for equilibration. The state variables, particles' trajectory, and density modes are recorded every 1000 steps equivalent to $2\tau$.

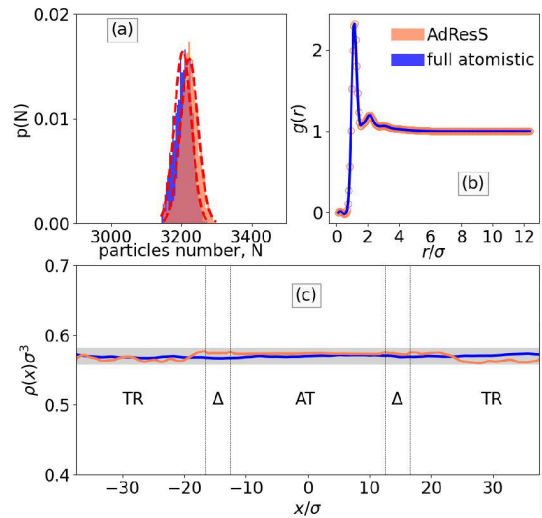

Figure 1. Comparison of the probability distribution of particles (a) in the region of interest (AT), radial distribution function (b), and density profile (c) for the AdResS and full atomistic simulation of reference in the equilibrium state of supercritical Argon at $\varrho^* = 0.57$ and $T^* = 2.5$.

**Validation**

For coupling AdResS to FHD, it is essential to ensure that the particle-based solver is properly modelling the fluid under study in different situations in and out of equilibrium. In the case of the thermal and hydrodynamic equilibrium where there is no pressure and temperature difference between different subregions of the particle subdomain, one expects to get the same equilibrium results for the AdResS and reference simulation. Apart from the flat density and temperature profiles, the same physical behaviour is expected. The quantities that describe the physical state of the system are the particles' probability distribution ($p(N)$) and the radial distribution function ($g(r)$) in the region of interest. The comparison of these properties at equilibrium for the Argon fluid at $\varrho^* = 0.57$ and $T^* = 2.5$ is shown in Fig.1.

While coupling FHD to AdResS, it is possible to have different densities and temperatures at the left and right sides of the particle subdomain, which turns the simulation into a non-equilibrium-like problem within the AdResS domain. Recently, Ebrahimi Viand et al. [4] performed AdResS simulations for non-equilibrium problems with different temperatures and densities at the left and right reservoirs. They studied the analogy of AdResS and Bergmann-Lebowitz model of open systems[10, 11] and performed AdResS simulation for the non-equilibrium problem by applying the thermodynamic forces for AdResS simulation of the equilibrium problems corresponding to each reservoir. In their analogy, once the AdResS system is in contact with two different reservoirs with different thermodynamic conditions, the combined effect translates to the combined thermodynamic forces for each side. In our case, we calculate two different thermodynamic forces for the left and right sides of the domain, at equilibrium conditions corresponding to each reservoir's thermodynamic condition. Next, we apply each of those thermodynamic forces to the left and right sides simultaneously, while setting the thermostat of the left and right sides according to the temperature of the reservoirs.

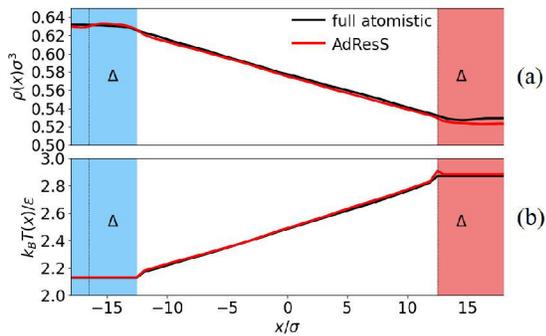

Figure 2. The profile of density (a) and temperature (b) for the non-equilibrium problem for AdResS and full atomistic simulation of reference where the left and right sides' reservoirs are at different thermodynamic state points with $\varrho^*_{left} = 0.63$, $T^*_{left} = 2.125$, $\varrho^*_{right} = 0.52$, and $T^*_{right} = 2.875$. The white region in the middle represents the AT region which is connected to the cold (blue) and hot (red) reservoirs.

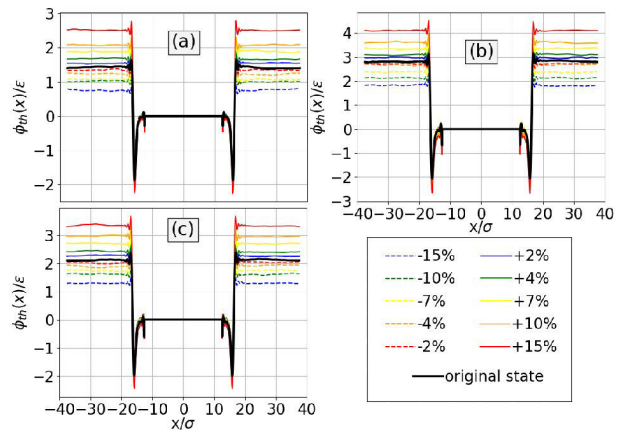

Figure 3. The potential of thermodynamic force for 3 different temperatures of $255[K]$(a), $345[K]$(b), and $300[K]$(c) in 11 different densities that have been calculated by iterative manner for AdResS simulations at equilibrium. The range of densities and temperatures covers $\pm\%15$ around the target state ($\varrho^* = 0.57$ and $T^* = 2.5$). The black solid line shows the potential of thermodynamic force for the original target state. The coloured solid and dashed lines represent the cases with a density higher and lower than the target state, respectively. In each set, densities increase in the order of colours such as blue, green, yellow, orange, and red.

Here, the most drastic scenario for the temperature gradient is considered where the cold and hot reservoirs are at the temperature of $T_{cold} = 255[K]$ ($T^*_{cold} = 2.125$) and $T_{hot} = 345[K]$ ($T^*_{hot} = 2.875$). Then, the density of the left and right sides' reservoirs is set in such a way that yields to $P(\varrho_{cold}, T_{cold}) = P(\varrho_{hot}, T_{hot})$. The results of the temperature and density gradient for the described non-equilibrium case are shown in Fig.2.

### Dictionary of thermodynamic forces

Performing the non-equilibrium simulation within AdResS[4] allows us to implement any non-equilibrium problem within a range by using a set of pre-calculated thermodynamic forces of the equilibrium situations with different reservoirs. This brings the idea of preparing a list of thermodynamic forces, called the "dictionary of thermodynamic forces", for a range of states and using the list to obtain the proper thermodynamic force for any other situation by interpolation. Thus, with the selected state of Argon, a set of AdResS simulations for three different temperatures and eleven different densities in the range of $\pm\%15$ around the target state are performed and the thermodynamic forces for the corresponding equilibrium cases are calculated. The potential of thermodynamic force for the abovementioned list of densities and temperatures is calculated and presented in Fig.3.

### FLUCTUATING HYDRODYNAMICS (FHD)

In this work, the full 3D description of the LLNS equation has been simplified to a 1D equation set:

$$\frac{\partial}{\partial t} \begin{pmatrix} \varrho \\ J \\ E \end{pmatrix} = -\frac{\partial}{\partial x} \begin{pmatrix} \varrho u \\ \varrho u^2 + P \\ (E+P)u \end{pmatrix} + \frac{\partial}{\partial x} \begin{pmatrix} 0 \\ \frac{4}{3}\eta \partial_x u \\ \frac{4}{3}\eta u \partial_x u + k \partial_x T \end{pmatrix} + \frac{\partial}{\partial x} \begin{pmatrix} 0 \\ s \\ q + us \end{pmatrix} \quad (3)$$

There are several discretization methods based on CFD schemes that are commonly used for the NS equations and could be extended to LLNS equations like MacCormack[12–14] and Piecewise Parabolic Method[13, 15, 16]. Here, we focus on the variance-preserving third-order Runge-Kutta scheme (RK3) as it shows to have more accuracy compared to the other mentioned schemes[13, 17]. The RK3 scheme can be written in the following three-stage form,

$$\begin{aligned}
\mathbf{U}_j^{n+1/3} &= \mathbf{U}_j^n + (\frac{\Delta t}{\Delta x})\Delta\mathcal{F}^n \\
\mathbf{U}_j^{n+2/3} &= \frac{3}{4}\mathbf{U}_j^n + \frac{1}{4}\mathbf{U}_j^{n+1/3} + \frac{1}{4}(\frac{\Delta t}{\Delta x})\Delta\mathcal{F}^{n+1/3} \quad (4)\\
\mathbf{U}_j^{n+1} &= \frac{1}{3}\mathbf{U}_j^n + \frac{2}{3}\mathbf{U}_j^{n+2/3} + \frac{2}{3}(\frac{\Delta t}{\Delta x})\Delta\mathcal{F}^{n+2/3}
\end{aligned}$$

where $\mathcal{F} = -\mathbf{F} + \mathbf{D} + \mathbf{S}$, $\Delta\mathcal{F} = \mathcal{F}_{j+1/2} - \mathcal{F}_{j-1/2}$, and values at $j \pm 1/2$ are a simple finite difference approximation of the cell centre values. The approximation of stochastic stress tensor ($s_{j+1/2}$) and heat flux ($q_{j+1/2}$) at the edge of the cells are[13]:

$$\begin{aligned}
q_{j+1/2}^n &= \sqrt{\frac{k_B}{\Delta t V_c}(k_{j+1}T_{j+1}^2 + k_j T_j^2)}\mathcal{R}(\mu,\sigma^2), \\
s_{j+1/2}^n &= \sqrt{\frac{4k_B}{3\Delta t V_c}(\eta_{j+1}T_{j+1} + \eta_j T_j)}\mathcal{R}(\mu,\sigma^2)
\end{aligned} \quad (5)$$

with $V_c$ being the volume of each continuum cell and $\mathcal{R}(\mu,\sigma^2)$ is independent Gaussian distributed random number with zero mean and unit variance. Bell and collaborators[13] showed that with the abovementioned stochastic fluxes, the flux's variance reduces to half of its original magnitude; thus, they suggested to use $\mathbf{S}_{new} = \mathbf{S}\sqrt{2}$ instead in all steps in Eq.4 and consequently change the definition of $\mathcal{F}$ to $\mathcal{F} = -\mathbf{F} + \mathbf{D} + \mathbf{S}\sqrt{2}$.

In a numeric simulation, the advective and diffusive stability criteria determine the maximum time step and its relation to space discretization size,

$$(|u|+c_s)\frac{\Delta t}{\Delta x} \leqslant 1, \quad max\left(\frac{4}{3}\frac{\overline{\eta}}{\overline{\varrho}}, \frac{\overline{k}}{\overline{\varrho}\,\overline{c_v}}\right)\frac{\Delta t}{\Delta x^2} \leqslant \frac{1}{2} \quad (6)$$

in which $c_s$ is the speed of sound and all over-line parameters are those values in the equilibrium state. Moreover, for the stability of the advection term, the discretization scheme is suitable for small cell's Reynolds numbers ($Re_c << 2$)[13]; this means that to have a small Reynolds number, $Re_c = \overline{\varrho u}\Delta x/\overline{\mu}$, one has to choose very small cell size. Accordingly, a small size for the space discretization is selected which restricts us to choose small time steps according to the stability inequalities in Eq.6. In all simulations, a periodic boundary condition is imposed.

## COUPLING ADRESS AND FHD

### Technical details

In all coupling simulations, the MD time step is set to $0.002\tau$ which is equivalent to $4.3[fs]$ and each AdResS simulation consists of $2.5 \times 10^6$ steps. The AT region of AdResS is designed to contain 10 continuum cells which means that the space discretization size ($dx$) for the continuum solver is $AT/10 = 2.5\sigma = 0.85[nm]$ while the whole domain size is $10AT = 250\sigma = 85[nm]$. The stability conditions restrict us to use a small time step for the continuum solver which in our case is set to $0.1[ps]$. The full MD simulation results are averaged over 20 independent NVE simulations for each test scenario after initializing by NVT simulations with the same MD parameters as the AsResS simulations but in the whole simulation domain with the size of $250\sigma \times 15\sigma \times 15\sigma$.

The FHD and MD solvers exchange information with each other every $10[ps]$. The continuum solver provides the thermodynamic state at the interface of the MD subdomain to calculate the proper thermodynamic force from the dictionary of thermodynamic forces (Fig.3). On the other hand, the MD solver provides the full atomistic information in the AT region and corresponding fluxes at the interface for the continuum subdomain. These interface fluxes include mass flux (particles mass crossing the interface), momentum flux (particles momentum crossing the interface), and energy flux (kinetic and interaction energy of particles on the opposite sides of the interface in AT and $\Delta$ region). However, due to the action of a thermostat and the unphysical nature of the tracer particles in the reservoir, it is required to check the calculated fluxes and adjust them to ensure they do not violate the conservation laws outside of the MD subdomain.

The coupling algorithm reads that after performing FHD simulation for the whole domain, the proper thermodynamic forces based on the thermodynamic state point resulting from FHD simulation at the interface are calculated for the left and right sides' reservoirs of AdResS. Then, the AdResS simulation is executed using those forces and the average quantities for $\varrho$, $u$, and $T$ in the AT region plus the abovementioned interface fluxes are calculated and passed to the FHD solver. The new values resulting from the MD simulation are replaced into the equivalent particle subdomain and its neighbour cells and the FHD simulation runs again. This simulation protocol will execute until getting satisfactory results.

There are several methods for deriving the equation of state for LJ which in general can be categorized into two groups: those with a theoretical basis and those with a purely empirical basis; each of them applies to some ranges of densities and temperatures[18]. The Modified Benedict-Webb-Rubin equation of state used in this work is the one used by Nicolas et al.[19] and explained in detail in Ref.[18]. In their derivation, the following relation for the pressure in reduced units is concluded:

$$P^* = \varrho^* T^* + \sum_{i=1}^{8} a_i \varrho^{*(i+1)} + F\sum_{i=1}^{6} b_i \varrho^{*(2i+1)} \quad (7)$$

where the coefficients $a_i$ and $b_i$ are functions of temperature and represented in Ref.[18]. In Eq.7, $F = exp(-\lambda \varrho^{*2})$ and $\lambda$ is an adjustable parameter.

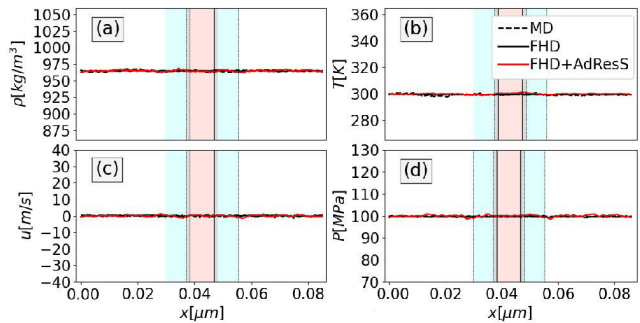

Figure 4. The profile of density (a), temperature (b), velocity (c), and pressure (d) for coupling AdResS to FHD with initial flat properties. The red shaded subregion shows the AT region of AdResS which is connected to the reservoir of non-interacting particles with blue shaded colour (TR) through a grey transition region ($\Delta$) where the vertical solid line between AT and $\Delta$ region represents the interface of MD and continuum subdomains. In all figures, the black solid and dashed lines show the result of the FHD and MD solvers, respectively, and the red solid line represents the result of coupling FHD to AdResS.

The thermodynamic properties (heat capacity, viscosity, and thermal conductivity) depend on the temperature and density and can be found at different state points of Argon in source data at the National Institute of Standards and Technology (NIST)[20]. In this work, we have found those values at 3 different temperatures and 11 different densities around the selected state within a range of $\pm\%15$ and interpolated for any other situations.

### Tests

Uniform density and temperature are applied as the initial condition to the AdResS-FHD coupling system. As there are no advective and diffusive forces in the domain (see Eq.3), it is expected that nothing should change during continuum simulation. After coupling to AdResS, some fluctuations will arise in the system because of the inevitable deviations of AdResS from the exact target state (less than 1.5%) and the natural fluid's fluctuations. In Fig.4, the profiles of thermodynamic and hydrodynamic properties of the system at some arbitrary times are shown. The initial condition with zero velocity is set with the described state of $\varrho = 964.82[kg/m^3]$, $T = 300[K]$, and $P = 100[MPa]$.

Next, similar to the initial step function presented in the main text, a sinusoidal wave for density and temperature is applied to the system, but with a non-uniform pressure profile. Here, the initial pressure function obeys a periodic behaviour; thus, the flow will have some oscillations until reaching the equilibrium state. This example shows how smoothly such oscillations will be handled with the developed coupling code while regenerating the results of the pure continuum and MD solvers. Here, the initial density and temperature oscillate around the target state of $\varrho = 964.82[kg/m^3]$ and $T = 300[K]$. One expects that these initial perturbations should resolve in the whole simulation domain over time due to the diffusive forces and the system should equilibrate at $\varrho = 964.82[kg/m^3]$ and $T = 300[K]$ after some oscillations. The results of this case are presented in Fig.5 over time. The profiles of density and temperature of AdResS and its reference full MD simulations are shown in Fig.6.

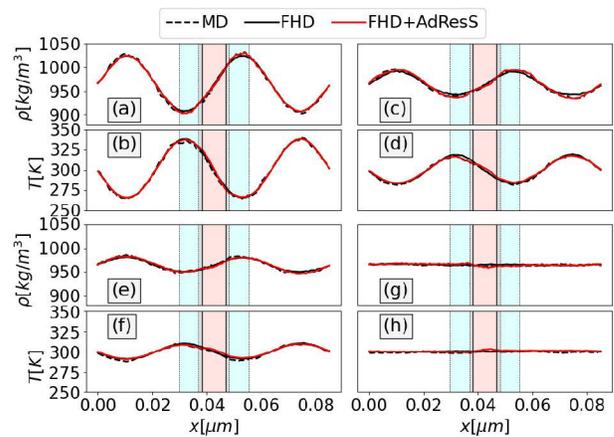

Figure 5. The profile of density (a, c, e, and g) and temperature (b, d, f, and h) of fluid with initial sinusoidal function for density and temperature overtime at $t = 0.0[ns]$ (a and b), $t = 0.25[ns]$ (c and d), $t = 0.45[ns]$ (e and f), and $t = 1.5[ns]$ (g and h). The initial density function is $\varrho(x) = 964.82(1 + 0.03 sin(4\pi x/l))$ and the temperature is $T(x) = 300(1 - 0.13 sin(4\pi x/l))$. The solid black line, dashed black line, and red line represent the results of FHD, MD, and coupling of AdResS to FHD, respectively.

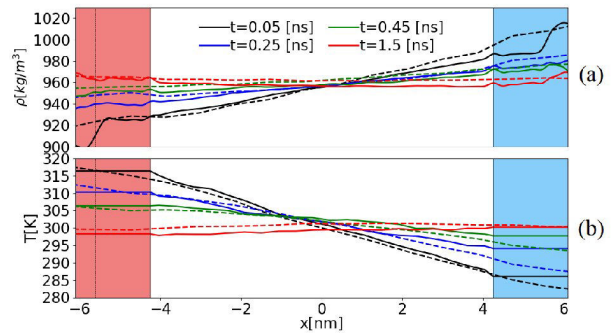

Figure 6. The profiles of density (a) and temperature (b) in the AdResS domain over time for the initial periodic density and temperature during coupling AdResS to FHD. The snapshots correspond to the times in Fig.5 and the order is illustrated in the labels. The regions with red and blue colours represent the initially hot and cold reservoirs, respectively.

Finally, to assess the capability of the AdResS-FHD coupling code for quasi one-dimensional problems, a new geometry with varying cross-sections is designed. Simi-lar geometries with varying cross-sections are relevant for bioengineering applications on micro and nanoscale for separation purposes, e.g. in acoustic wave microfluidic devices [21, 22]. The cross-section of the new geometry linearly increases in the continuum domain on the left and symmetrically decreases on the right side while having a constant value at the middle part corresponding to the AT region. Applying such conditions to the continuum solver needs some corrections in the discretization formula as the cell's cross-section area is a function of the length, $x$. This requires the addition of a factor of $A(x)$ to all terms in Eq.3 which leads to the same factor in all terms in the discretization algorithm of Eq.4. Furthermore, applying such a change in the geometry will result in a source term of $P \partial A / \partial x$ to the momentum equation of the Landau-Lifshitz Navier-Stokes equation set (Eq.3) [23]. The cross-section area in the middle is similar to the previous cases with $A_{middle} = (15\sigma)^2$ and its value at the left and right borders of the domain is $A_{left} = A_{right} = (12.25\sigma)^2$. The results of the AdResS-FHD simulation are shown for density and temperature in Fig.7 over time and the new solver with coupling AdResS to FHD is calculating the evolution of the system correctly.

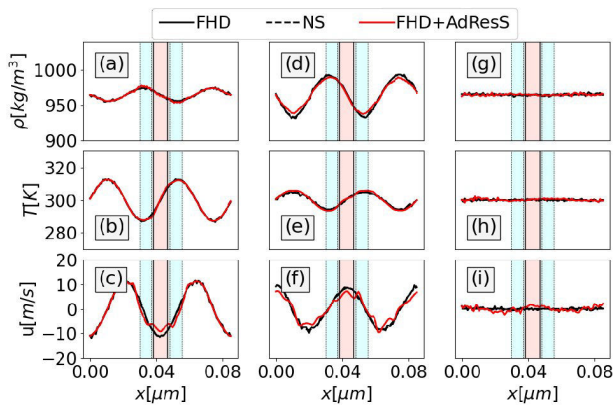

Figure 7. The evolution of density (a, d, and g), temperature (b, e, and h), and flow velocity (c, f, and i) for the case with varying cross-section where it increases and decreases linearly in the left and right continuum subdomains, respectively, while being constant in the middle at $t = 0.01[ns]$ (a, b, and c), $t = 0.1[ns]$ (d, e, and f), and $t = 10[ns]$ (g, h, and i). The coupling simulation is started with an arbitrary initial condition which is set to an initial uniform density and sinusoidal temperature with an oscillation amplitude of $15[K]$ around the target state.